# Active switching in metamaterials using polarization control of light

Hua Xu and Byoung Seung Ham[*]

*Center for Photon Information Processing, School of Electrical Engineering,
and the Graduate School of Information and Communications,
Inha University, Incheon 402-751, Republic of Korea*

## Abstract

We demonstrate on-demand control of localized surface plasmons in metamaterials by means of incident light polarization. An asymmetric mode, selectively excited by *s*-polarized light, interfere destructively with a bright element, thereby allowing the incident light to propagate at a fairly low loss, corresponding to electromagnetically induced transparency (EIT) in an atomic system. In contrast, a symmetric mode, excited by *p*-polarized light, directly couples with the incident light, which is analogous to the switch-off of EIT. The light polarization-dependent excitation of asymmetric and symmetric plasmon modes holds potential for active switching applications of plasmon hybridization.

---

[*] bham@inha.ac.kr; Corresponding author

Electromagnetically induced transparency (EIT), which directly results from destructive quantum interference between alternative pathways, renders an opaque medium transparent over a narrow spectral range, resulting in a dramatic variation of refractive index. The EIT phenomenon has been applied to quantum nonlinear optics such as slow light, enhanced optical nonlinearities, ultrafast switching, and optical delay lines [1, 2]. Recently, a mimic of EIT in classical systems has drawn much attention to surface plasmon studies to reduce dissipation in metamaterials [3, 4], sensors [5], compact delay lines, and optical buffers [6]. In particular, plasmon-induced transparency [3, 7-9] is of great interest for merging plasmonics and metamaterials, with potential for magnetic and negative index metamaterials [10, 11] at optical frequencies as well as integrated plasmon-optical circuits.

EIT in an atomic system requires two independent laser beams: one for probe and the other for pump [12]. In contrast, EIT in metamaterials requires an incident light field only, creating two plasmonic modes. Compared with EIT in an atomic system, the realization of plasmonic EIT in metamaterials is passive [13-15]. Here we propose an active scheme of plasmon-induced transparency for dynamic plasmon mode switching in plasmon hybridization. In this scheme, symmetric and asymmetric plasmon modes are designed to be degenerate for a certain wavelength. Depending on incident light polarization, the degeneracy can be broken for a dark mode. Accordingly, the speed of dynamic plasmon mode control is determined by the speed of light polarization.

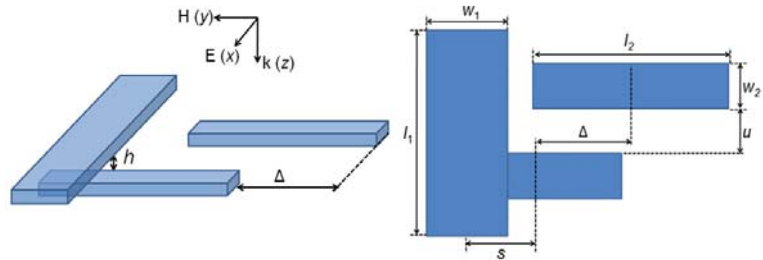

Fig. 1. (Color online) Schematic of the unit cell at oblique and top views, and the configuration of incident light. The unit cell is comprised of three metal strips with the geometrical parameters: $l_1$ = 136 nm, $W_1$ = 50 nm, $l_2$ = 120 nm, $W_2$ = 30 nm, $u$ = 30 nm.

The thickness of each strip is 20 nm. They are stacked vertically with a distance of $h = 30$ nm, where the spacer is treated as air for simplicity without any loss of generality. The displacement of the upper and the lower strips is taken to $s = 35$ nm. The longitudinal shift of the two metal strips is denoted by $\Delta$, which is a key factor in controlling the plasmonic modes.

The structure of a unit cell is depicted in Fig. 1, where one silver strip is located in the upper layer, and two parallel silver strips are in the lower layer. In particular, the two lower layer strips have a longitudinal shift $\Delta$, which is the critical factor in controlling the plasmon hybridization [14]. A plane wave of light is incident on the unit cell perpendicularly. Both *s* (electric component) and *p* polarizations (magnetic component) parallel to the upper strip are considered. We use the finite-difference time-domain (FDTD) method [16-18] for numerical calculations. The permittivity of silver is described by a Drude formula [19].

Due to the plasmon hybridization, a compound structure consisting of several elementary components exhibits two pronounced resonances (*i.e.,* lifted degeneracies) [13]. One resonance at a shorter wavelength is associated with a symmetric or superradiant plasmonic mode, ascribed to electric resonance; the other at a longer wavelength is associated with an asymmetric or subradiant plasmonic mode, attributed to magnetic resonance [14]. However, the spectral positions of these two modes can be controlled by a longitudinal shift $\Delta$ [14, 15]. Unlike the superradiant (bright) element, the subradiant element hardly couples with the incident light due to its vanishing dipole moment [3, 20]. In order to mimic EIT, the design of the subradiant (dark) element is of great importance using two parallel metal strips with a longitudinal shift $\Delta$ as shown in Fig. 1.

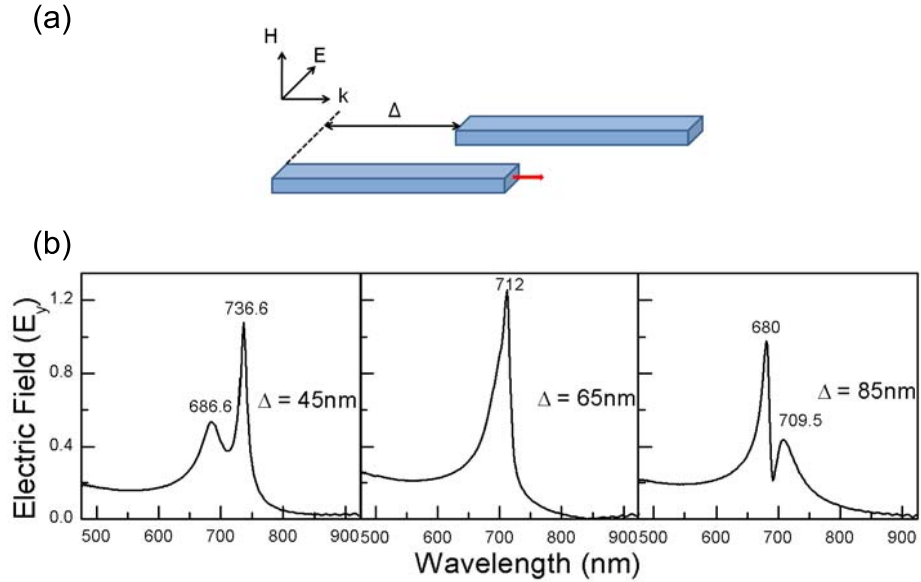

Fig. 2. (Color online) Excitation of the individual subradiant element. (a) Incident wave is parallel to the metal pair and its magnetic component is perpendicular to the plane of the pair. The longitudinal shift Δ is varied from 45 to 85 nm. (b) Corresponding spectral responses are investigated by an *Ey* probe placed 10 nm away from the center of the end facet of one of the metal pair.

In order to clearly illustrate the two resonance modes, the spectral responses are presented in a near-field zone for various shift Δ, as shown in Fig. 2. A far-field zone represents a weak coupling of the asymmetric mode with the incident light. Here the plane wave of light enters horizontally at a grazing angle, since the asymmetric mode can be activated by a vertical component of the magnetic field [3, 21]. In the numerical calculations, two pronounced resonances exist for the case of Δ = 45 nm and Δ = 85 nm. In Fig. 2(b), for Δ = 45 nm, one peak at 686.6 nm represents the symmetric mode; the other peak at 736.6 nm represents the asymmetric mode. In contrast, for the case of Δ = 85 nm, the modes are swapped. In the middle at Δ=65 nm, two modes cross over at the same point. This spectral variation is due to the evolution of Coulomb forces in the metal pair resulting from the interaction of charges located at the strip ends [14]. Interestingly, the two resonances nearly overlap at the same wavelength, 712 nm for Δ=65 nm, illustrating that the symmetric and the asymmetric modes are degenerate at the same wavelength. We now use this degeneracy for dynamic switching of plasmon modes by selection of incident light polarization.

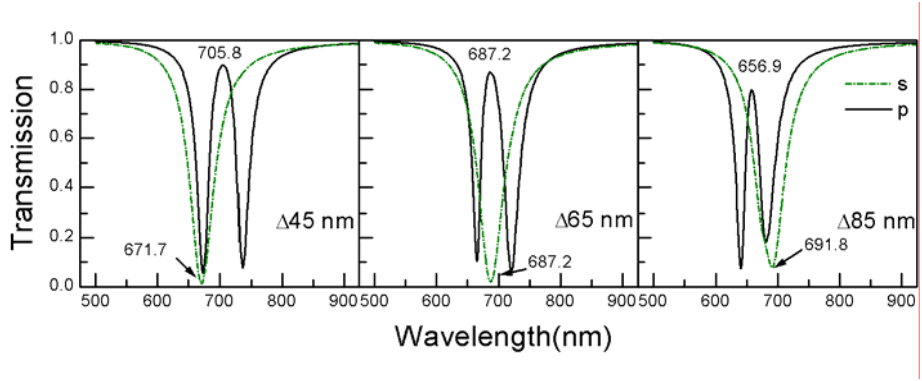

Fig. 3. (Color online) Transmission spectra for either *s* or *p* polarization with Δ = 45, 65, and 85 nm.

Figure 3 shows numerical calculations for Fig. 1 for Δ = 45, 65, and 85 nm. When Δ = 65 nm, the absorption peak of the symmetric mode (dashed line) for *p* polarization of the incident light exactly coincides with the transmission peak (transparency) at line center of the asymmetric mode for *s* polarization (solid line). If the *s* polarized light is incident vertically, then only the asymmetric mode is selectively excited at the metal pair by the upper strip through the near-field interactions, resulting in the EIT-like peak at line center ($\lambda$ = 687.2 nm), which is explained as a dipole-quadrupole coupling [7,8]. For Δ = 45nm, the asymmetric mode is located on the right side of the symmetric mode. For Δ = 85nm, the positions of the asymmetric and symmetric modes are reversed. This phenomenon coincides with the results of Fig. 2. Compared with the resonant wavelength in the near-field zone [see Fig. 2(b)], the EIT-like peaks in Fig. 3 are all blue shifted, which could be attributed to the interaction of the upper strip with the lower pair and/or the retardation effect from the near-field to the far-field zones [22]. For the *p* polarized incident light, the EIT-like spectrum is totally wiped out and turns out into absorption (see the dashed-dot line). Obviously, the plasmonic EIT can be manipulated by selecting the polarization of light without an extra incident field.

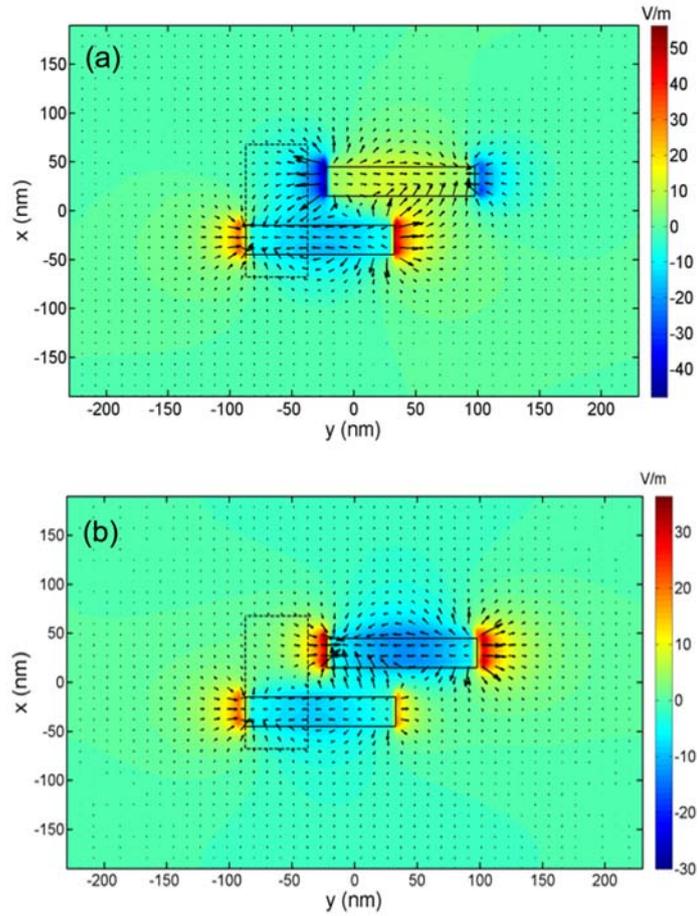

Fig. 4. (Color online) (a) Distribution of electric field $E_y$ for the incidence with $s$ polarization at the wavelength of plasmon-induced transparency, 687.2 nm. The electric field vector overlays $E_y$ for better visualization. (b) Distribution of $E_y$ and the vector of the electric field for the incidence with $p$ polarization at the wavelength of the transmission dip, 687.2 nm.

To speculate the physical origin of mode control using the light polarization in Fig. 3, we take a closer look at the distribution of the induced electric fields for $s$ and $p$ polarizations at the wavelength of 687.2 nm. Figure 4(a) shows that electric field $E_y$ is mainly localized at the ends of the metal pair but shows a moderate concentration between the pair strips due to Δ. Figure 4(a) presents an asymmetric mode for $s$ polarization. The color at the ends of the strips shows the asymmetric feature. Although the asymmetric mode cannot be directly excited by the external optical field at normal incidence due to the lack of a vertical H component, it is accessible by the near-field interactions, where the external incidence with $s$ polarization resonantly excites the upper metal strip first, and then interacts with the lower strips in the form of dipole-quadrupole coupling. If the polarization of the incident light changes to *p*, the electric field illustrates a distinct behavior differing from the case of *s* polarization. As shown in Fig. 4(b), the distribution of the electric field at the end of the pair strips is the same as in the symmetric mode. Because the symmetric mode is able to couple with the external incidence directly, the near-field interaction does not take effect. As a result, the vanishing transmission appears when the light is vertically incident on the surface with *p* polarization and excites the symmetric mode in the lower strips. Therefore, a plasmonic EIT switching can be realized by choosing the polarization of the incident light. Accordingly, the approach taken to alter the polarization determines the method needed to tune the EIT-like effect. The plasmonic EIT could be switched on and off by either an electric or optical method. For instance, if the proposed structure is integrated with a Pockels cell, where the Pockels cell is used for light polarization switching based on induced birefringence by an applied voltage [23], electro-optic switching can be achieved. If the proposed structure is integrated with a multiple quantum well, where an optically addressed polarization can be switched based on the near-resonant excitation of a spin-polarized excitons [24], all-optical switching can be realized.

In conclusion, we investigated on-demand plasmonic switching using polarization selection of incident light. The degeneracy of symmetric and asymmetric modes in the plasmon hybridization is achieved by elaborately adjusting the longitudinal shift of the metal pair for a resonant wavelength. Based on this, we proposed a scheme by which the plasmonic EIT can be switched on and off, depending on the polarization selection of the incident light, since degeneracy holds for wavelength, not for polarization in such a structure. The proposed idea could be integrated with a Pockels cell or multiple quantum wells to switch the EIT-like effect.

This work was supported by the CRI program (No. 2010-0000690) of the Korean Ministry of Education, Science and Technology via National Research Foundation.

[1] J. P. Marangos, J. Mod. Opt. **45**, 471 (1998).

[2] M. Fleischhauer, A. Imamoglu, and J. P. Marangos, Rev. Mod. Phys. **77**, 633 (2005).

[3] S. Zhang, D. A. Genov, Y. Wang, M. Liu, and X. Zhang, Phys. Rev. Lett. **101**, 047401 (2008).

[4] P. Tassin, L. Zhang, Th. Koschny, E. N. Economou, and C. M. Soukoulis, Phys. Rev. Lett. **102**, 053901 (2009).

[5] N. Liu, T. Weiss, M. Mesch, L. Langguth, U. Eigenthaler, M. Hirscher, C. Sonnichsen, and H. Giessen, Nano Lett. **10**, 1103 (2010).

[6] R. W. Boyd and D. J. Gauthier, Nature (London) **441**, 701 (2006).

[7] N. Liu, L. Langguth, T. Weiss, J. Kastel, M. Fleischhauer, T. Pfau, and H. Giessen, Nat. Mater. **8**, 758 (2009).

[8] R. D. Kekatpure, E. S. Barnard, W. Cai, and M. L. Brongersma, Phys. Rev. Lett. **104**, 243902 (2010).

[9] K. Aydin, I. M. Pryce, and H. A. Atwater, Opt. Express **18**, 13407 (2010).

[10] S. Linden, C. Enkrich, M. Wegener, J. Zhou, T. Koschny, and C. M. Soukoulis, Science **306**, 1351 (2004).

[11] J. Valentine, S. Zhang, T. Zentgraf, E. Ulin-Avila, D. A. Genov, G. Bartal, and X. Zhang, Nature (London) **455**, 376 (2008).

[12] S. E. Harris, Phys. Today **50**, 36 (1997).

[13] E. Prodan, C. Radloff, N. J. Halas, and P. Nordlander, Science **302**, 419 (2003).

[14] B. Kante, S. N. Burokur, A. Sellier, A. de Lustrac, and J.-M. Lourtioz, Phys. Rev. B **79**, 075121 (2009).

[15] A. Christ, Y. Ekinci, H. H. Solak, N. A. Gippius, S. G. Tikhodeev, and O. J. F. Martin, Phys. Rev. B **76**, 201405(R) (2007).

[16] A. Taflove and S. C. Hagness, *Computational electrodynamics: The finite-difference time-domain method* (Artech: Norwood, MA, 2000).

[17] A. Farjadpour, D. Roundy, A. Rodriguesz, M. Ibanescu, P. Bermel, J. D. Joannopoulos, S. G. Johnson, and G. Burr, Opt. Lett. **31**, 2972 (2006).

[18] A. F. Oskooi, D. Roundy, M. Ibanescu, P. Bermel, J. D. Joannopoulos, and S. G. Johnson, Comput. Phys. Commun. **181**, 687 (2010).

[19] See supplementary material at [URL will be inserted by AIP] for the simulation details and the identification of symmetric and asymmetric modes from the distribution of electric field when the subradiant mode is excited individually.

[20] M. I. Stockman, S. V. Faleev, and D. J. Bergman, Phys. Rev. Lett. **87**, 167401 (2001).

[21] S. A. Maier, Nat. Mater. **8**, 699 (2009).

[22] G. W. Bryant, F. J. Garcia, and J. Aizpurua, Nano Lett. **8**, 631 (2008).

[23] E. Hecht, *Optics*, 4th ed. (Addison Wesley, San Francisco, 2002).

[24] E. J. Gansen, K. Jarasiunas, and A. L. Smirl, Appl. Phys. Lett. 80, 971 (2002).